\documentclass[preprint,aps]{revtex4}

\usepackage{graphicx}

\begin{document}

\title{Electronic Evidence of Temperature-Induced Lifshitz Transition and Topological Nature in ZrTe$_5$}

 \author{Yan Zhang$^{1,\sharp}$, Chenlu Wang$^{1,\sharp}$, Li Yu$^{1,\sharp}$, Guodong Liu$^{1,*}$, Aiji Liang$^{1}$, Jianwei Huang$^{1}$, Simin Nie$^{1}$, Yuxiao Zhang$^{1}$, Bing Shen$^{1}$, Jing Liu$^{1}$, Hongming Weng$^{1,2}$, Lingxiao Zhao$^{1}$, Genfu Chen$^{1,2}$, Xiaowen Jia$^{3}$, Cheng Hu$^{1}$, Ying Ding$^{1}$, Shaolong He$^{1}$, Lin Zhao$^{1}$, Fengfeng Zhang$^{4}$, Shenjin Zhang$^{4}$, Feng Yang$^{4}$, Zhimin Wang$^{4}$, Qinjun Peng$^{4}$, Xi Dai$^{1,2}$, Zhong Fang$^{1,2}$, Zuyan Xu$^{4}$, Chuangtian Chen$^{4}$ and X. J. Zhou$^{1,2,*}$}

\affiliation{
\\$^{1}$Beijing National Laboratory for Condensed Matter Physics, Institute of Physics, Chinese Academy of Sciences, Beijing 100190, China.
\\$^{2}$Collaborative Innovation Center of Quantum Matter, Beijing 100871, China.
\\$^{3}$Military Transportation University, Tianjin 300161, China.
\\$^{4}$Technical Institute of Physics and Chemistry, Chinese Academy of Sciences, Beijing 100190, China.
}

\date{February 11, 2016}

\maketitle

{\bf The topological materials, including topological insulators\cite{HasanReveiw,QiReview,AndoReview}, three-dimensional Dirac semimetals\cite{BurkovDirac,YoungDirac,WangNa3Bi,WangCd3As2,LiuNa3Bi,NeupaneCd3As2,LiuCd3As2,BorisenkoCd3As2,YiCd3As2,SYX_SCIENCE_2015} and three-dimensional Weyl semimetals\cite{WanWeyl,BurkovWeyl,HuangTaAs,WengTaAs,BQLvTaAsNP,XuNbAS,LvPRX,XuTaAs}, have attracted much attention recently for their unique electronic structure, spin texture and peculiar physical properties. While three-dimensional topological insulators are becoming abundant, two-dimensional topological insulators remain rare\cite{Bernevig,MKonig}, particularly in natural materials\cite{AndoReview}. ZrTe$_5$ has host a long-standing puzzle on its anomalous transport properties manifested by its anomalous resistivity peak; its underlying origin remains elusive\cite{Furuseth,Weiting,OS_JPSJ_1980,JT_SSC_1982}. Lately, ZrTe$_5$ has ignited renewed interest because it is predicted that single-layer ZrTe$_5$ is a two-dimensional topological insulator and there is possibly a topological phase transition in bulk ZrTe$_5$\cite{HW_PRX_2014}. However, the topological nature of ZrTe$_5$ is under debate as some experiments point to its being a three-dimensional Dirac semimetal\cite{QL_ARXIV_2014,RYC_PRB_2015,RYC_PRL_2015} or  a quasi-two-dimensional Dirac semimetal\cite{XY_ARXIV_2015}. Here we report high-resolution laser-based angle-resolved photoemission measurements on the electronic structure and its detailed temperature evolution of ZrTe$_5$. The electronic property of ZrTe$_5$ is dominated by two branches of nearly-linear-dispersion bands at the Brillouin zone center. These two bands are separated by an energy gap that decreases with decreasing temperature but persists down to the lowest temperature we measured ($\sim$2 K). The overall electronic structure exhibits a dramatic temperature dependence; it evolves from a {\it p}-type semimetal with a hole-like Fermi pocket at high temperature, to a semiconductor around $\sim$135 K where its resistivity exhibits a peak, to an {\it n}-type semimetal with an electron-like Fermi pocket at low temperature. These results indicate a clear electronic evidence of the temperature-induced Lifshitz transition in ZrTe$_5$. They provide a natural understanding on the underlying origin of the resistivity anomaly at $\sim$135 K and its associated reversal of the charge carrier type. Our observations also provide key information on deciphering the topological nature of ZrTe$_5$ and possible temperature-induced topological phase transition.}

The transition metal pentatellurides like ZrTe$_5$ and HfTe$_5$\cite{Furuseth} have attracted considerable interest since the last 70s because they exhibit unusual transport properties characterized by a strong resistivity peak\cite{Weiting,OS_JPSJ_1980}  accompanied by a sign reversal of the Hall coefficient and thermopower across the peak temperature\cite{Izumi_1982,JT_SSC_1982,DNM_JPCM_2004}.  The origin of such transport property anomaly has been a subject of a long-time debate but remains unclear, with explored possibilities of structural phase transtion\cite{OS_JPSJ_1980,DFJ_PRB_1981}, formation of a charge/spin density wave\cite{OS_JPSJ_1982}, polaronic behavior\cite{MR_PRB_1999}, a semimetal-semiconductor transition\cite{DNM_JPCM_2004}, or temperature-induced band shift\cite{GM_PRL_2015}. With the emergence of topological materials, including topological insulators\cite{HasanReveiw,QiReview,AndoReview}, three-dimensional Dirac semimetals\cite{BurkovDirac,YoungDirac,WangNa3Bi,WangCd3As2,LiuNa3Bi,NeupaneCd3As2,LiuCd3As2,BorisenkoCd3As2,YiCd3As2,SYX_SCIENCE_2015} and three-dimensional Weyl semimetals\cite{WanWeyl,BurkovWeyl,HuangTaAs,WengTaAs,BQLvTaAsNP,XuNbAS,LvPRX,XuTaAs}, ZrTe$_5$ and HfTe$_5$ have ignited renewed interest as a candidate of a novel topological material. While many three-dimensional topological insulators have been predicted and discovered, two-dimensional topological insulators, also known as quantum spin Hall insulators that can support topologically-protected helical edge modes inside a bulk insulating gap and lead to dissipationless transport, are rare\cite{Bernevig,MKonig}, especially in natural compounds\cite{AndoReview}.  It was predicted that single-layer ZrTe$_5$ is a promising candidate of a two-dimensional topological insulator with a large bulk band gap, while bulk ZrTe$_5$ may host a possibility of realizing a temperature-driven topological phase transition between the weak and strong topological insulators\cite{HW_PRX_2014}. However, a number of recent experiments on ZrTe$_5$ point to its being a three-dimensional Dirac semimetal\cite{QL_ARXIV_2014,RYC_PRB_2015,RYC_PRL_2015} or a quasi-two-dimensional Dirac semimetal\cite{XY_ARXIV_2015}. Added to the interest of ZrTe$_5$ is the recent observation of superconductivity under high pressure\cite{YHZ_ARXIV_2015}. Direct investigation on the electronic structure of ZrTe$_5$ is highly-desired in understanding the electronic origin of the transport property anomaly, and in uncovering the exact nature of the topological state in ZrTe$_5$.

Angle-resolved photoemission spectroscopy (ARPES)\cite{ARPESReview} can provide direct information on the electronic structure of materials in addressing the above-mentioned prominent issues in ZrTe$_5$. However, high-resolution comprehensive ARPES measurements on ZrTe$_5$ are still lacking\cite{DNM_JPCM_2004,QL_ARXIV_2014,GM_PRL_2015}. In this paper, we present detailed ARPES measurements on ZrTe$_5$ with unprecedented energy and momentum resolutions, by taking advantage of our latest generation laser-based ARPES system which can cover two-dimensional momentum space simultaneously.  We find that the electronic property of ZrTe$_5$ is dominated by two branches of bands with nearly-linear dispersion, a valence band and a conduction band, at the Brillouin zone center. There is an energy gap that separates the two branches of bands; the gap decreases with decreasing temperature and persists to a very low temperature ($\sim$2 K). The overall electronic structure exhibits a dramatic temperature dependence that shifts down with decreasing temperature. This results in an evolution from a {\it p}-type semimetal with a hole-like Fermi pocket at high temperature above $\sim$135 K, to a semiconductor around $\sim$135 K where its resistivity shows a peak, to an {\it n}-type semimetal with an electron-like Fermi pocket at low temperature.  These results constitute strong electronic evidence of a temperature-induced Lifshitz transition in ZrTe$_5$.  They provide a natural understanding on the underlying origin of the resistivity anomaly at $\sim$135 K and its associated sign reversal of Hall coefficient and thermopower. Our observations also provide key information to reveal the nature of the topological state in ZrTe$_5$.

ZrTe$_5$ is a layered compound which crystallizes in the orthorhombic crystal structure\cite{Furuseth}. The ZrTe$_5$ crystal is constructed from layer-stacking along the {\it b}-axis of the ZrTe$_5$ sheets. The two-dimensional ZrTe$_5$ sheet (Fig. 1a) features a trigonal prismatic ZrTe$_6$ chains running along the {\it a}-axis that are linked together along the {\it c}-axis via zigzag chains of Te atoms. Each ZrTe$_5$ sheet is nominally charge neutral, and the distance between the adjacent two ZrTe$_5$ sheets along the {\it b}-axis is quite large ($\sim$7.25 $\AA$), giving rise to a weak interlayer coupling\cite{HW_PRX_2014}.  Fig. 1b shows the projected surface Brillouin zone for ZrTe$_5$ with the high symmetry momentum points labeled in a standard way. The resistivity-temperature data of our ZrTe$_5$ sample (Fig. 1c) exhibits a prominent peak at $\sim$135 K. This is consistent with the typical ZrTe$_5$ samples reported before\cite{Weiting,OS_JPSJ_1980,DFJ_PRB_1981,JT_SSC_1982,OS_JPSJ_1982,MR_PRB_1999,GK_PRB_1985,MI_JPC_1987,DNM_JPCM_2004,YHZ_ARXIV_2015,XY_ARXIV_2015,GM_PRL_2015} (see Methods for the sample details).

The ARPES data are taken with our new laser-based system equipped with the latest-generation time-of-flight analyzer. It not only has high energy and momentum resolutions, but also has a new capability of covering two-dimensional momentum space simultaneously (see Methods for experimental details).  Fig. 1d-g shows the measured constant energy contours of ZrTe$_5$ at 195 K, which represent the spectral intensity distribution at the Fermi level (Fig. 1d) and at a few binding energies of 100 meV (Fig. 1e), 200 meV (Fig. 1f) and 300 meV (Fig. 1g). The corresponding band structures are shown in Fig. 1h-k measured along several typical momentum cuts as indicated in Fig. 1g.  We have checked the effect of the polarization geometry on the measured results.  The measured constant energy contours and band structures are similar under two typical {\it s} and {\it p} polarization geometries, although the spectral weight distribution in the momentum space shows a clear polarization dependence (see Fig. S1 in Supplementary Materials).

The constant energy contour at the Fermi level shows a tiny spot at the Brillouin zone center, $\Gamma$ point (Fig. 1d). With increasing binding energy, it grows in area and evolves into a warped rectangle at high binding energies of 200 meV (Fig. 1f) and 300 meV (Fig. 1g).  This is consistent with the observation of hole-like bands below the Fermi level as seen in Fig. 1h-k. These hole-like bands show nearly-linear dispersions over a wide energy range. They also exhibit a strong anisotropy: the bands are steeper along the horizontal or vertical high-symmetry directions (Fig. 1h and 1j for cuts 1 and 3 in Fig. 1g) with a Fermi velocity of $\sim$4 eV$\cdot$${\AA}$ than those along the momentum cuts 2 (Fig. 1i) and 4 (Fig. 1k) with a Fermi velocity of $\sim$2.3 eV$\cdot$${\AA}$.  Detailed lineshape analysis of these hole-like bands at $\Gamma$ indicates that they are broad; the corresponding  photoemission spectra (energy distribution curves, EDCs)  are not composed of a single sharp peak or two sharp peaks, but represent a spectral continuum encompassed by two edges (see Fig. S2 in Supplementary Materials).

To understand the measured electronic structure, we have performed band structure calculations of ZrTe$_5$ (see Methods for details of the band structure calculations). The calculated band structures (Fig. S3 in Supplementary Materials) and constant energy contours (Fig. S4 in Supplementary Materials) are basically consistent with the previous reports\cite{MW_PRB_1982,HW_PRX_2014}.  The evolution of the constant energy contours near $\Gamma$ point with the increasing binding energy (Fig. 1d-g) from a spot to a small circle to a warped rectangle is in good agreement with the calculated results (Fig. S4 in Supplementary Materials), considering a chemical potential difference in the comparison. Band structure calculations on bulk ZrTe$_5$ give only one sharp hole-like band near $\Gamma$ in this case that is apparently not consistent with our measurements. The slab calculations (Fig. S3b), on the other hand, provide a good explanation on this behavior. This broad band feature can be understood reasonably well by the finite k$_z$ effect, consistent with the results of the slab band calculations (Fig. S3b and S3d).

We have carried out detailed ARPES measurements on ZrTe$_5$ at different temperatures in order to understand the origin of the anomalous transport properties and examine on possible temperature-induced topological phase transition in ZrTe$_5$.  Fig. 2 shows the temperature dependence of the energy bands measured along two high-symmetry momentum cuts. The corresponding temperature dependence of the constant energy contours is shown in Fig. 3. These results are highly reproducible by measuring on the same sample with cycles of cooling down and warming up (see Fig. S5 in Supplementary Materials), by cleaving samples at high temperatures and low temperatures, and by measuring on many different samples. The band images (Fig. 2a-b) are obtained by dividing the original data with their corresponding Fermi distribution functions, making it possible to reveal features above the Fermi level at relatively high temperatures.

From Fig. 2a and 2b, it is clear that the band structure near $\Gamma$ consists of two branches, one is the upper branch (UB) at lower binding energy or above the Fermi level that corresponds to the electron-like conduction band, while the other is the lower branch (LB) at high binding energy that corresponds to the hole-like valence band. In between the UB and LB bands, there is a region with strongly-suppressed spectral weight that we call it valley region.  The bands at $\Gamma$ point show a strong temperature dependence; the overall band structure shifts down to high binding energy with decreasing temperature. Specifically, the LB band touches the Fermi level at high temperature (255 K in Fig. 2a and 2b); it shifts down with decreasing temperature and becomes well below the Fermi level at low temperatures. On the other hand, the UB band is well above the Fermi level at high temperature (255K in Fig. 2a and 2b), moves down with decreasing temperature, and crosses the Fermi level at very low temperature (e.g., 35 K and 2 K in Fig. 2a and 2b). Such a band shift with temperature is consistent with those reported before\cite{DNM_JPCM_2004,GM_PRL_2015}.

Although the overall band structure shifts down with decreasing temperature, we find that it is not a rigid band shift.  To keep track on the temperature evolution of the bands in a more quantitative way, we plot the photoemission spectra (EDCs) at $\Gamma$ point measured at different temperatures in Fig. 2c.  The EDCs consist of signals from the LB band and the UB band, with a valley separating in between them. The LB band is visible in the entire measured temperature range which is encompassed by a lower-binding-energy upper edge  and high-binding-energy lower edge. The UB band is invisible above E$_F$ at high temperatures (255 K), moves down with decreasing temperature, and part of it becomes visible at low temperatures.  Fig. 2d shows the energy positions of the upper edge of the LB band, lower edge of the LB band, and the center of the valley bottom. The two edges of the LB band and the valley center show a similar trend of temperature dependence; the overall energy shift from 255 K to 2 K is on the order of $\sim$70 meV.  A close examination indicates that the LB band gets wider with decreasing temperature  as seen from the energy difference between the upper and lower edges of the LB band (Fig. 2e). On the other hand, the U-shaped valley  becomes steeper with decreasing temperature, seen from the energy difference between the upper edge of the LB band and the valley center (Fig. 2e). These indicate that the band shift with temperature is not a strictly rigid band shift.  We notice that 135 K is a characteristic temperature where the valley center crosses the Fermi level (Fig. 2c and 2d).  We also note that the energy shift of bands near $\Gamma$ with temperature is faster in the high temperature region of 165$\sim$255 K than that in the lower temperature 2$\sim$105 K region (Fig. 2d).

The strong suppression of the spectral weight in the valley region is indicative of an gap opening between the LB band and the UB band (Fig. 2a and 2b). As seen in Fig. 2c, at high temperatures, the valley bottom in EDCs shows a flat region with an intensity close to zero when taking the signal background into account. This suggests a true gap opening at high temperatures. We estimate the gap size in two ways. In the first method, we choose the high binding energy region at 0.30$\sim$0.35 eV in EDCs which is very weak  and slightly above zero intensity as the background line. The line intersects with the valley bottom at two points and the distance between these two points are taken as an estimation of the gap size. Compared with zero intensity background, this method slightly over-estimates the gap size, but it provides an objective and direct way in extracting the gap value and the difference is small compared to that obtained from zero intensity background.  The energy gap obtained in this way is large at high temperatures ($\sim$40 meV for 255 K), and gets smaller with decreasing temperature (Fig. 2f). At very low temperatures, such as 35 K and 2 K,  the spectral weight near the valley bottom increases that is above the zero intensity. In this case, there is no more true gap opening so we cannot determine the gap size using the same criterion as used at high temperatures. However, we believe there remains a gap opening near the valley region because the spectral weight is still strongly suppressed, a situation similar to the pseudogap observed in high temperature cuprate superconductors\cite{ARPESReview}. It is also possible that, besides a true gap opening, there is a new in-gap state produced by other mechanism that fills the gap region and grows with decreasing temperature. If there is no gap opening and a Dirac cone-like structure is formed, one would expect to see a peak at the Dirac point that is not consistent with the present result. This leads us to take another way to estimate the gap size.  If we assume a Dirac cone structure as a reference point that has zero gap, the gap opening would split the LB and UB bands and causes a spectral weight suppression at the gapped valley region. In this case, the distance between the LB and UB bands can be used as a measure of the gap size. Assuming the LB and UB bands show similar temperature-dependent splitting from the valley center, the gap size can be estimated as twice the distance between the position of the upper edge of the LB band and the position of the valley center, as shown in Fig. 2f (red  empty circles). In this case, the gap size still decreases with decreasing temperature, but it persists in the entire temperature range we have measured.

The temperature dependence of the Fermi surface provides a clear evidence of a temperature-induced Lifshitz transition in ZrTe$_5$.  Fig. 3 shows the temperature evolution of the constant energy contours at the Fermi level (Fig. 3a) and at a binding energy of 100 meV (Fig. 3b) for ZrTe$_5$.  The corresponding momentum distribution curves (MDCs) across the $\Gamma$ point at different temperatures are shown in Fig. 3c-d and Fig. 3e-f for the two cases, which provide information on the temperature evolution of the pocket size (Fig. 3g) and the spectral weight at the Fermi level (Fig. 3h).  At a high temperature like 255 K, there is a tiny hole-like pocket at $\Gamma$ (left-most panel in Fig. 3a), consistent with a hole-like band touching the Fermi level in Fig. 2a and 2b. With decreasing temperature, the hole-like pocket shrinks in size, becomes invisible at 135 K, then emerges again below 135 K.  Its size increases with further decreasing of temperature and it becomes an electron-like pocket as seen from bands in Fig. 2a and 2b. Therefore, there is a clear Lifshitz transition that occurs across $\sim$135 K where the Fermi surface topology transforms from a hole-like pocket at high temperature to an electron-like pocket at low temperature. For the constant energy contours at a binding energy of 100 meV (Fig. 3b), the pocket size at the $\Gamma$ point keeps shrinking with decreasing temperature, as seen from the distance of two peaks in MDCs (Fig. 3e-f and Fig. 3i). As seen in Fig. 2, the band at 100 meV binding energy represents solely the LB valence band over the entire temperature range we measured. It is natural that the pocket size shrinks when the hole-like band moves to high binding energy with decreasing temperature.

The identification of a temperature-induced Lifshtz transition provides direct information to naturally explain the origin of the transport property anomalies in ZrTe$_5$.  Because there is an energy gap between the LB band and UB band, and the bands shift with temperature (Fig. 2), it is clear that at high temperature, when the Fermi surface is a hole-like pocket, it is a {\it p}-type semimetal. With decreasing temperature, the hole-like valence band sinks down to below the Fermi level and the Fermi level lies in the gapped region, ZrTe$_5$ enters a semiconducting state. At $\sim$135 K, the Fermi level is close to the center of the energy gap. With further decreasing of temperature, the UB conduction band moves downwards to touch the Fermi level. The Fermi surface becomes an electron-like pocket and ZrTe$_5$ changes into an {\it n}-type semimetal at low temperature.  Therefore, ZrTe$_5$ undergoes a transition from a {\it p}-type semimetal at high temperature, to a semiconductor in a narrow temperature region around 135 K, to an {\it n}-type semi-metal at low temperature. The temperature evolution of the spectral weight at the Fermi level (Fig. 3h) is consistent with this picture. We note that, around 135 K, no other bands except for those near $\Gamma$ cross the Fermi level, as seen from the band structure calculations (Fig. S2 and S3 in Supplementary Materials) and the measurements at extremely low temperature (Fig. 4).  This provides a natural explanation on the resistivity maximum at $\sim$ 135 K in the transport measurement. It also accounts for the sign reversal of the Hall coefficient and thermopower because it corresponds to charge carrier change from {\it p}-type at high temperature to {\it n}-type at low temperature across $\sim$135 K.

Our present results pose a new issue on the origin of the chemical potential variation with temperature in ZrTe$_5$, i.e., how temperature can induce a dramatic chemical potential shift in ZrTe$_5$ without obvious charge carrier doping. In terms of usual understanding, our temperature-dependent results indicate that extra holes are present above $\sim$135 K, no carriers present in the semiconducting state around 135 K, and extra electrons appear below $\sim$135 K. Without external doping mechanism, these seem to violate the charge balance in the temperature variation process. We believe the measured temperature-dependent behaviors represent intrinsic properties of ZrTe$_5$, not artifact caused by extrinsic factors like trivial absorption/desorption process. First,  the measurements were performed in ultra-high vacuum and in different ARPES systems; the results are reproducible.  Second, the results are consistent with other ARPES results on similar ZrTe$_5$ sample with similar resistivity anomaly temperature\cite{DNM_JPCM_2004,GM_PRL_2015}. Third, our ARPES measurements identified a Lifshitz transition at $\sim$135 K that corresponds to the resistivity anomaly temperature; it is hard to believe this is coincidental. Fourth, most importantly, our ARPES results are consistent with the bulk Hall coefficient and thermopower measurements that indicate a charge carrier change from hole-like at high temperature to electron-like at low temperature across $\sim$135 K\cite{Izumi_1982,JT_SSC_1982,DNM_JPCM_2004}.  Such an unusual temperature-dependent chemical potential shift was considered to be a characteristic of a semiconductor\cite{DNM_JPCM_2004}. It may also be related to charge carrier localization/delocalization upon changing temperature. Considering similar temperature dependence of the band gap (Fig. 2f) and the lattice parameter {\it b} (Fig. 3j), it is also interesting to investigate whether the variation of the interlayer interaction with temperature might give rise to such an unusual electronic state transition. While we do not have a clear answer on the origin yet, the dramatic charge carrier change with temperature may hide some deep physical mechanism that calls for further investigations.

Our detailed electronic structure measurements of ZrTe$_5$ provide direct information to examine on the nature of its possible topological state, i.e., whether it is a topological insulator\cite{HW_PRX_2014} or a quasi-two-dimensional\cite{XY_ARXIV_2015} or three-dimensional Dirac semimetal\cite{QL_ARXIV_2014,RYC_PRB_2015,RYC_PRL_2015}. If ZrTe$_5$ is a quasi-two-dimensional semimetal, one would observe a Dirac cone in the {\it ac}-plane that we have measured. The absence of such a Dirac cone in our measurements clearly rules out such a scenario.  For a similar reason, the gap opening between the LB valence band and the UB conduction band, especially its variation with temperature,  and its persistence to the lowest temperature we measured (Fig. 2c and 2f), are not compatible with the three-dimensional Dirac semimetal picture for ZrTe$_5$. Our electronic structure measurements are in good agreement with the transport measurements that show a resistivity anomaly and a reversal between electron and hole-like charge carriers.  One may argue whether the gap opening we observed is due to k$_z$ effect because, in principle, the three-dimensional Dirac cone can only be seen at particular k$_z$ values.  The gradual temperature-dependent change of the relative position between the UB and LB bands near $\Gamma$ point provides strong evidence on the gap opening, irrespective of k$_z$ location, which is not consistent with a three-dimensional Dirac cone picture. We note that the inconsistency on the nature of topological state in ZrTe$_5$  may be due to sample difference. The ZrTe$_5$ we measured here has a resistivity anomaly at $\sim$135 K that is consistent with most of the samples reported before\cite{Weiting,OS_JPSJ_1980,DFJ_PRB_1981,JT_SSC_1982,OS_JPSJ_1982,MR_PRB_1999,GK_PRB_1985,MI_JPC_1987,DNM_JPCM_2004,YHZ_ARXIV_2015,XY_ARXIV_2015,GM_PRL_2015}. On the other hand, for the reports of
three-dimensional Dirac semimetal\cite{QL_ARXIV_2014,RYC_PRB_2015,RYC_PRL_2015}, the ZrTe$_5$ samples were grown by a different method and exhibit a resistivity anomaly at a much lower temperature, $\sim$60 K. While our ARPES results are consistent with previous reports\cite{DNM_JPCM_2004,GM_PRL_2015} on the ZrTe$_5$ samples with similar resistivity anomaly temperature $\sim$135 K, they are quite different from those on the ZrTe$_5$ sample with a $\sim$60 K resistivity anomaly\cite{QL_ARXIV_2014}.

Band structure calculations predict that single-layer ZrTe$_5$ is a two-dimensional topological insulator\cite{HW_PRX_2014}. Our ARPES results are hard to provide a definitive answer on this issue because the ZrTe$_5$ sample we measured is a bulk, not a single-layer. Moreover, the two-dimensional topological insulator has edge states that ARPES is usually not sensitive to detect. However, we observed some new features that are possibly compatible with the ZrTe$_5$ edge states.  For some cleaved thin ZrTe$_5$ samples, or cleaved thin samples that have many one-dimensional thread structures caused by experiencing many cycles of temperature variation (see Fig. S6b in Supplementary Materials), we often observe obvious extra bands growing on top of the original bulk bands (Fig. S6g).  In the corresponding constant energy contours (Fig. S6c-e), these extra bands form quasi-one-dimensional intensity streaks running along the k$_y$ direction, which is perpendicular to the one-dimensional ZrTe$_6$ chain direction that runs along {\it a}-axis.  We note that the signal intensity of these extra bands varies between samples but its observation is quite common.  These extra bands are not present in the band structure calculations of the bulk ZrTe$_5$ (Fig. S3).  Their observations are compatible with the possibility of coming from sample boundary or edge states. Recent scanning tunneling microscope (STM) experiments on ZrTe$_5$ reported observation of edge states on ZrTe$_5$ surface\cite{XBLi,RWu}. Further efforts are needed to pin down on the origin of these new features we observed from ARPES.

We finally examine on possible temperature-induced topological phase transition expected in ZrTe$_5$\cite{HW_PRX_2014}. Band structure calculations\cite{HW_PRX_2014} indicate that the topological property of ZrTe$_5$ is sensitive to the interlayer coupling between the adjacent ZrTe$_5$ layers and bulk ZrTe$_5$ is sitting very close to the border between the weak and strong topological insulators. When the interlayer coupling is weak, bulk ZrTe$_5$ acts as a three-dimensional weak topological insulator. When the interlayer coupling gets strong, bulk ZrTe$_5$ can be transformed into a three-dimensional strong topological insulator like typical Bi$_2$Se$_3$ and Bi$_2$Te$_3$ three-dimensional topological insulators\cite{HJZhang,ZHasan,YLChen}. In this case, topological surface state is expected to appear, as shown in Fig. S3c and S3e (see Supplementary Materials).  With decreasing temperature, the lattice constant {\it b} decreases (Fig. 3j),  the gap between the LB band and the UB band decreases; these are consistent with the enhancement of the interlayer coupling  as expected\cite{HW_PRX_2014}.

To check whether the gap between the LB and UB bands near $\Gamma$ is closed and possible emergence of the in-gap topological surface state at low temperature, we carried out ARPES measurement on ZrTe$_5$ to a very low temperature, $\sim$2 K (Fig. 4). At such an extremely low temperature, in addition to the electronic states at $\Gamma$, we observe another band (denoted as $\beta$ band) that crosses the Fermi level and forms four tiny electron pockets near ($\pm$0.14, $\pm$0.22) $\AA^{-1}$ points (Fig. 4a). This is consistent with the band structure calculations (Fig. S4d in Supplementary Materials). From the band structure (Fig. 4c) and the corresponding photoemission spectra (EDCs, Fig. 4e), the band bottom of this $\beta$ band is only $\sim$5 meV below the Fermi level; its bottom is above the bottom of the UB band at $\Gamma$ point by $\sim$15 meV.  The observation of this $\beta$ electron pocket, together with the $\alpha$ electron pocket at the $\Gamma$ point, is consistent with the quantum oscillation measurements where two electron pockets are detected\cite{GK_PRB_1985,MI_JPC_1987}. The appearance of the $\beta$ electron pockets will also enhance electrical conductivity at low temperature.

The topological transition from a weak to a strong three-dimensional insulators asks that the LB and UB bands at $\Gamma$ point move close to each other with decreasing temperature, touch and merge together, and then open a new gap after band inversion.  Our ARPES results (Fig. 2) indeed show a trend of gap closing with decreasing temperature, but we believe the gap persists over the entire temperature range we measured. One might argue that below 65 K, the filling up of the spectral weight at the valley bottom could be due to appearance of a new surface state that emerges in the new gapped region after the original gap is closed and the band inversion is realized. This possibility cannot be fully ruled out. To check on this possibility, we carefully searched for the possible signature of the topological surface state that would be expected inside the gap region near $\Gamma$ point at $\sim$2 K.  As seen from Fig. 4c and 4f, no signature of the extra surface state can be resolved. Therefore, our results suggest that, till the lowest temperature we measured ($\sim$2 K), the gap between the LB band and the UB  band near $\Gamma$ remains open and no band inversion has been realized in the measured temperature range. The lattice constant shrinkage of {\it b}-axis with temperature (Fig. 3j) by 0.3$\%$ is not enough for driving the topological phase transition to occur\cite{HW_PRX_2014}. Our present study indicates a clear trend of such a transition; further enhancement of the interlayer coupling is needed to materialize such a topological phase transition.

In summary, we have carried out comprehensive high resolution ARPES measurements on ZrTe$_5$. We have uncovered strong electronic evidence of a temperature-induced Lifshitz transition in ZrTe$_5$. The sample undergoes a {\it p}-type-semimetal to a semiconductor, to an {\it n}-type-semimetal transition with decreasing temperature. This evolution provides a natural explanation on the long-standing issue of the origin of the resistivity anomaly in ZrTe$_5$. It also indicates clearly that ZrTe$_5$ with a resistive anomaly at $\sim$135 K is not a three-dimensional Dirac semimetal. We found electronic signatures that may be related to the edge states associated with two-dimensional topological insulator. With decreasing temperature, there is a trend of the transition from a weak topological insulator to a strong topological insulator, but the strong topological insulator has not been realized at the lowest temperature we measured ($\sim$2 K). Further enhancement of interlayer coupling, either by external high pressure or internal chemical pressure, may facilitate to the realization of such a transition.

\vspace{3mm}
{\bf Methods}

High quality single crystal samples of ZrTe$_5$  were grown by the chemical vapour transport method with iodine as transport agent. The unique resistivity peak sits at a temperature of $\sim$135 K (Fig. 1c), in good agreement with the typical values reported in most of previous literatures\cite{Weiting,OS_JPSJ_1980,DFJ_PRB_1981,JT_SSC_1982,OS_JPSJ_1982,MR_PRB_1999,GK_PRB_1985,MI_JPC_1987,DNM_JPCM_2004,YHZ_ARXIV_2015,XY_ARXIV_2015,GM_PRL_2015}.

The ARPES measurements were performed at our new laser-based system equipped with the 6.994 eV vacuum-ultra-violet (VUV) laser and the time-of-flight electron energy analyzer (ARToF 10K by Omicron Scienta)\cite{YXZ_ARXIV_2015}. This latest-generation ARPES system is capable of measuring photoelectrons covering two-dimensional momentum space (k$_x$, k$_y$) simultaneously. Measurements were performed using both {\it s}- and {\it p}-polarization geometries with a laser repetition rate of 400 KHz. The overall energy resolution was set at 1-5 meV, and the angular resolution was 0.1$^o$. All the samples were measured in ultrahigh vacuum with a base pressure better than 5$\times$10$^{-11}$ mbar. The samples for temperature-dependent experiments were cleaved {\it in situ} at either 35 K or 255 K and measured at temperatures ranging from 35-255 K.  In order to avoid the hydrogen contamination on the sample surface at low temperature, we deliberately stay away from 20$\sim$33 K range near the hydrogen boiling temperature.

The band structure calculations are performed with the projector augmented wave (PAW) method\cite{blochl1994,kresse1999} implemented in Vienna {\it ab initio} simulation package (VASP)\cite{kresse1996_1,kresse1996_2}. The cut-off energy for the plane wave expansion is set to 450 eV. The generalized gradient approximation (GGA) of Perdew-Burke-Ernzerhof type\cite{Perdew1996} is used to deal with the exchange and correlation potential. The k-point sampling grid in the self-consistent process is 13$\times$13$\times$7. SOC is included as a second vibrational step using scalar relativistic eigenfunctions as the bases after the initial calculation is achieved in the self-consistent iterations. Maximally localized Wannier functions (MLWFs) for the p orbitals of Te atoms have been constructed by using the WANNIER90 code\cite{marzari1997,souza2001,mostofi2008}. A slab tight-binding model has been constructed to calculate the surface states by using the Wannier functions.

\vspace{3mm}

$^{\sharp}$These people contributed equally to the present work.\\

$^{*}$Corresponding author: gdliu\_arpes@iphy.ac.cn, XJZhou@aphy.iphy.ac.cn.

\vspace{3mm}

\noindent {\bf Acknowledgement}\\
This work is supported by the National Science Foundation of China (11574367), the 973 project of the Ministry of Science and Technology of China (2013CB921700, 2013CB921904 and 2015CB921300)£¬ and the Strategic Priority Research Program (B) of the Chinese Academy of Sciences (Grant No. XDB07020300).

\vspace{3mm}

\noindent {\bf Author Contributions}\\
Y.Z., C.L.W. and L.Y. contribute equally to this work. X.J.Z., G.D.L., Y.Z., C.L.W., L.Y., A.J.L. and J.W.H. proposed and designed the research. L.X.Z. and G.F.C. contributed in sample growth. S.M.N., H.W.M., X.D. and Z.F. contributed in the band structure calculations. Y.Z., C.L.W., L.Y., G.D.L., A.J.L., J.W.H., Y.X.Z., B.S., J.L., X.W.J., C.H., Y.D., S.L.H., L.Z., F.F.Z., S.J.Z., F.Y., Z.M.W., Q.J.P., Z.Y.X., C.T.C. and X.J.Z. contributed to the development and maintenance of the Laser-ARTOF system and related software development. Y.Z., C.L.W., J.W.H. and A.J.L. carried out the ARPES experiment.  Y.Z., C.L.W., L.Y., G.D.L., A.J.L. and X.J.Z. analyzed the data. X.J.Z., G.D.L. and Y.Z. wrote the paper with C.L.W., L.Y., A.J.L., S.M.N. and H.M.W.. All authors participated in discussion and comment on the paper.\\

\noindent{\bf Additional information}\\
Supplementary information is available in the online version of the paper.
Correspondence and requests for materials should be addressed to G.D.L. or X.J.Z.



\newpage

\begin{figure*}[tbp]
\begin{center}
\includegraphics[width=1.0\columnwidth,angle=0]{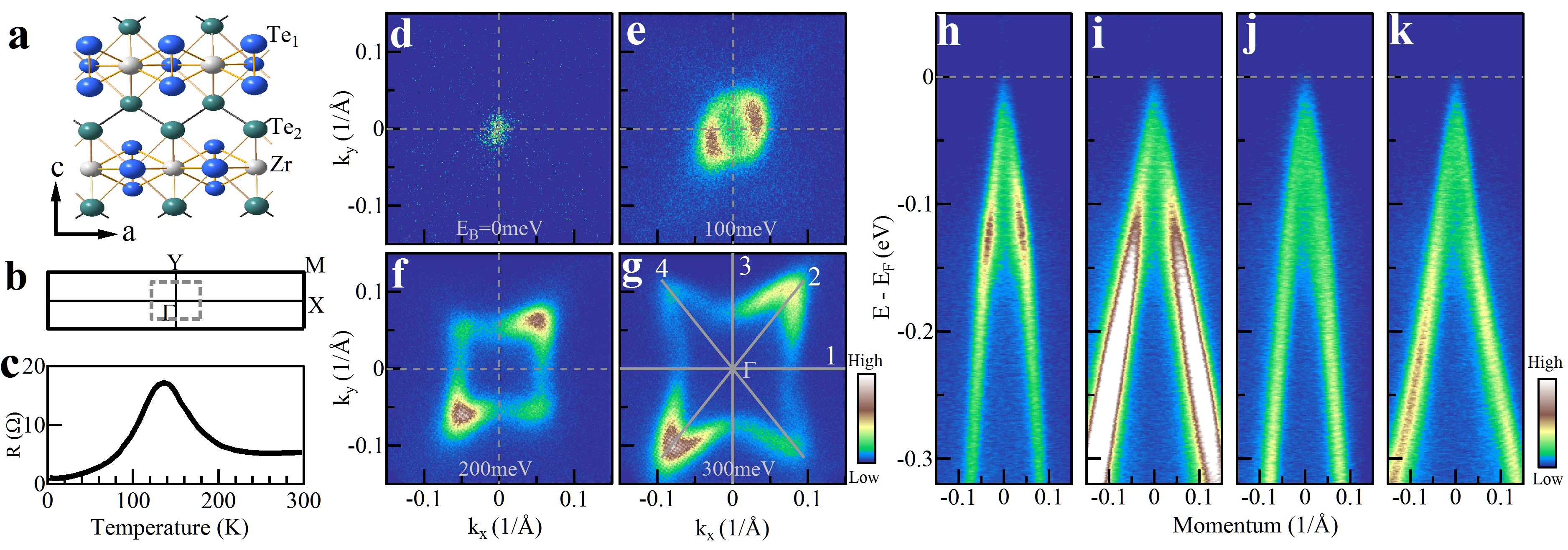}
\end{center}

\caption{{\bf Fermi surface and band structure of ZrTe$_5$ measured at 195 K.} (a) Top view of the bulk crystal structure ({\it ac} plane) of the ZrTe$_5$ sheet. The blue and green spheres represent Te atoms and the grey ones represent Zr atoms. ZrTe$_5$ crystal is constructed from stacking of the ZrTe$_5$ sheets along the {\it b}-axis (perpendicular to the {\it ac}-plane). (b) Surface Brillouin zone corresponding to {\it ac}-plane. High symmetry points are indicated. The central dashed-line square indicates the measured momentum space covered for by our ARPES mapping in (d-g). (c) Temperature dependence of resistivity for our ZrTe$_5$ single crystal samples; there is a clear resistivity peak at $\sim$135 K. (d-g) Constant energy contours of ZrTe$_5$ at different binding energies of 0, 100, 200 and 300 meV, respectively. The spectral intensity is integrated within 10 meV with respect to each binding energy.  The measurement geometry is set under {\it s} polarization. (h-k) Band structures measured along typical cuts 1, 2, 3 and 4, respectively.  The location of the momentum cuts is shown in (g) by thick grey lines.}

\end{figure*}

\begin{figure*}[tbp]
\begin{center}
\includegraphics[width=1.0\columnwidth,angle=0]{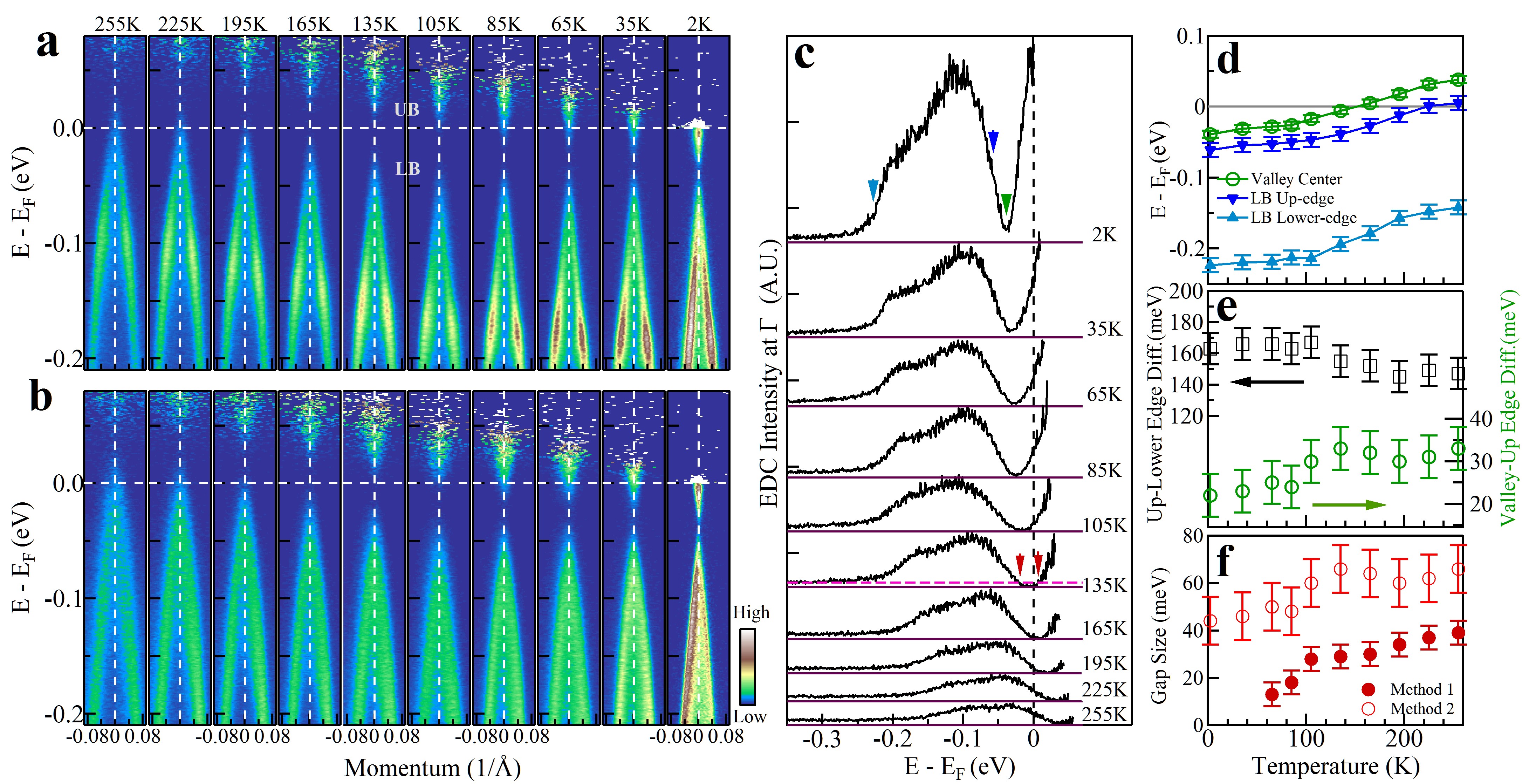}
\end{center}

\caption{{\bf Temperature evolution of the band structures in ZrTe$_5$.} (a-b) Temperature-dependent band structures measured along ${\Gamma}$-X (cut 1 in Fig. 1g) and ${\Gamma}$-Y directions (cut 3 in Fig. 1g). The corresponding Fermi distribution functions are divided out  in order to reveal features above the Fermi level.  (c) EDCs at $\Gamma$ point at different temperatures. For clarity, the EDCs are offset along vertical axis, with zero intensity represented by the horizontal purple lines. The EDCs consist of signals from the LB valence band and the UB conduction band, with a valley separating in between them with its center marked by a green triangle as for the 2 K EDC. The LB band is encompassed by a lower-binding-energy edge (marked by a deep blue arrow as for the 2 K EDC) and high-binding-energy edge (marked by a light blue arrow as for the 2 K EDC). (d) shows the energy positions of the high-binding-energy edge (light blue triangles) and lower-binding-energy edge (deep blue triangles) of the LB band, together with the center of the valley bottom (green empty circles) at different temperatures. (e) Energy difference between the two edges of the LB band (black empty squares) and between the lower-binding-energy edge of the LB band and the center of the valley bottom (green empty circles) at different temperatures.  (f) Energy gap size at different temperatures estimated from two methods. In method 1, we take the low intensity region of EDCs at high binding energy 0.30$\sim$0.35 eV as the background (horizontal dashed pink line for 135 K EDC) that intersects with the valley bottom at two points (as marked by two red arrows for 135 K EDC). The gap size (solid red circles) is estimated from the distance between these two points. In method 2, the gap size (empty red circles) is estimated as twice the energy difference between the valley bottom center and the upper edge of the LB band.
}
\end{figure*}

\begin{figure*}[tbp]
\begin{center}
\includegraphics[width=1.0\columnwidth,angle=0]{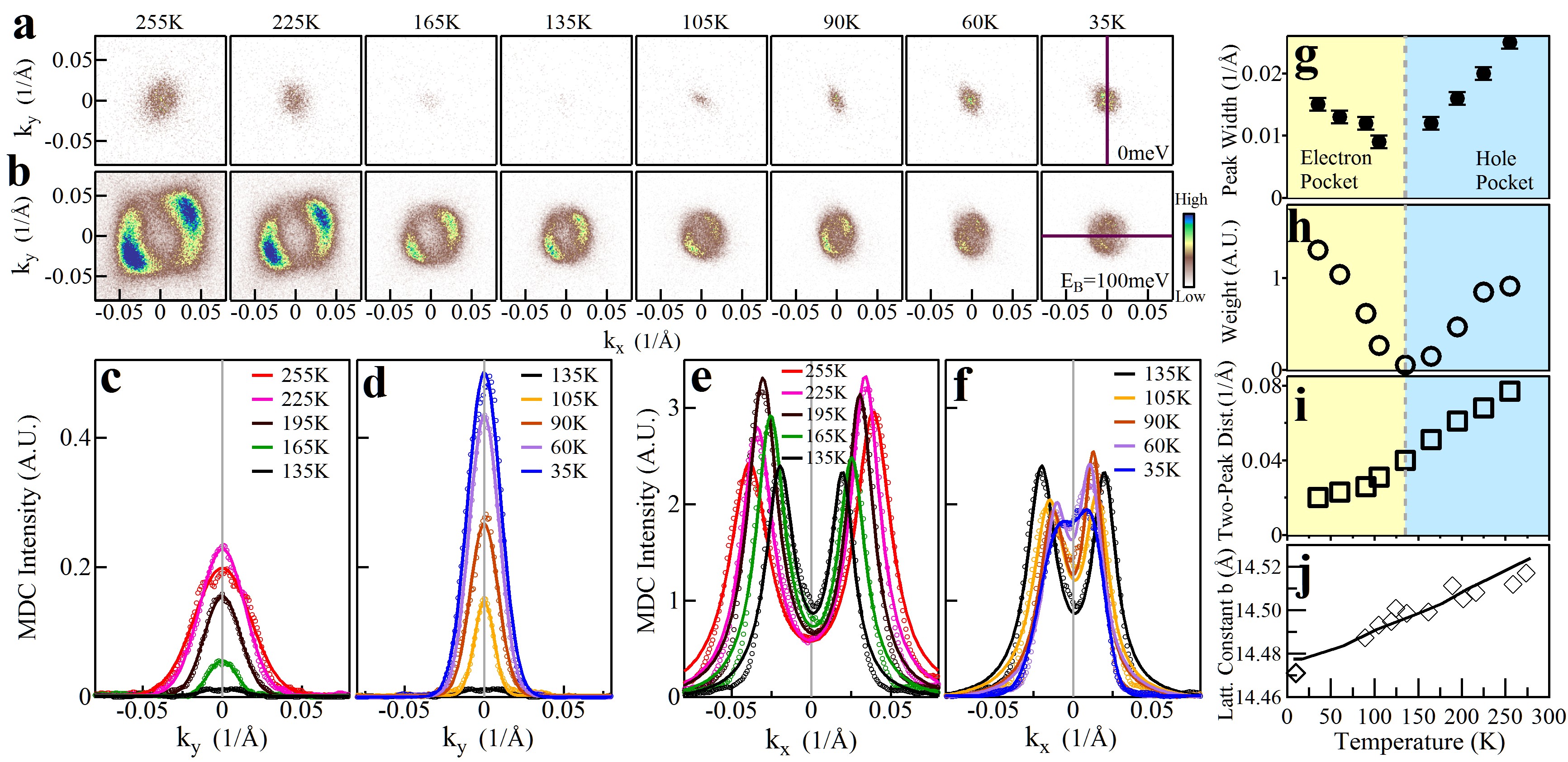}
\end{center}

\caption{{\bf Temperature-induced Lifshitz transition in ZrTe$_5$.} (a) Fermi surface evolution with temperature when ZrTe$_5$ is cooled down from 255 K to 35 K. (b) Corresponding constant energy contour evolution with temperature at a binding energy of 100 meV. (c-d) Momentum distribution curves (MDCs) at the Fermi level (E$_F$) measured along the vertical momentum cut (${\Gamma}$-Y direction) as indicated in the 35 K panel of (a) at different temperatures (open circles). To improve data statistics, the MDCs are obtained by integrating within $\pm$ 5 meV energy window with respect to the Fermi level. Because multiple-peak features are not resolved, we fitted the MDCs by a Gaussian to estimate the pocket size and signal intensity. The fitted MDC width and spectral weight are shown in (g) and (h), respectively.   (e-f) MDCs at a binding energy of 100 meV measured along the horizontal momentum cut (${\Gamma}$-X direction) as indicated in the 35 K panel of (b) at different temperatures (open circles). The MDCs are obtained by integrating within $\pm$ 5meV energy window at the binding energy of 100 meV. Here the MDCs show two clear peaks that are approximated by two Lorentzians or Gaussians. The distance between the two peaks is shown in (i) that is related to the area of the constant energy contours in (b). (g) Temperature dependence of the MDC width (FWHM) extracted from (c) and (d).  (h) Temperature dependence of the MDC weight, the integrated area of MDCs, extracted from (c) and (d).  (i) Temperature dependence of the two-peak distance of MDCs in (e) and (f).  (j) Temperature dependence of the lattice constant {\it b} with the measured data (black diamonds) and the fitted line (black line) which is related to the inter-layer spacing. Adopted from Ref.\cite{HF_SSC_1986}.}
\end{figure*}

\begin{figure*}[tbp]
\begin{center}
\includegraphics[width=1.0\columnwidth,angle=0]{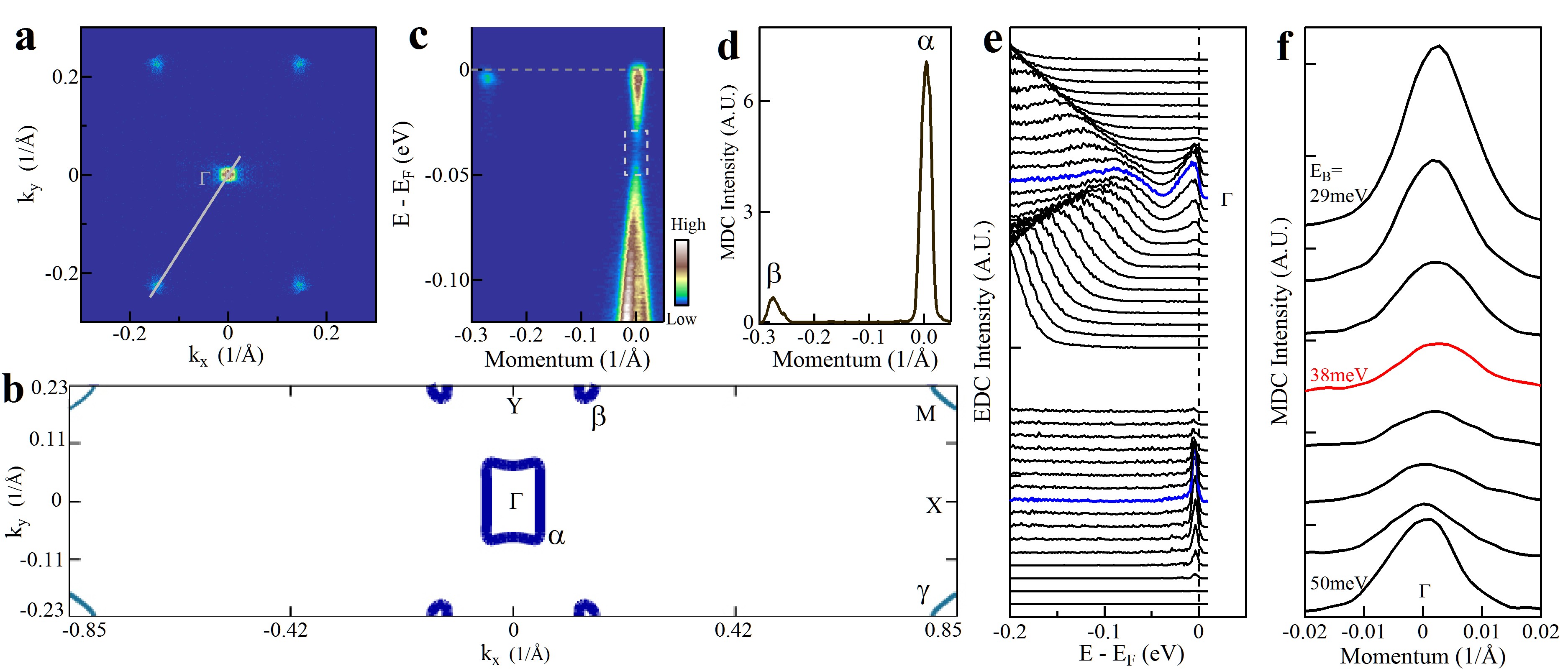}
\end{center}

\caption{{\bf Electronic structure of ZrTe$_5$ measured at a lower temperature of $\sim$2 K.} (a) Measured Fermi surface around the Brillouin zone center. In addition to the electron pocket at $\Gamma$, four more electron pockets are observed near ($\pm$0.14, $\pm$0.22) $\AA^{-1}$ points.  (b) Calculated Fermi surface under the weak topological insulator solution with a k$_z$=0.2 $\pi$/b. It shows a good agreement with the measured Fermi surface in (a). The $\gamma$ pocket near M point is beyond the momentum coverage of our 6.994 eV laser.  (c) A typical band structure measured along the momentum cut marked in (a) that crosses two electron pockets. (d) Corresponding MDC for the band structure in (c) at the Fermi level showing two peak structures. (e) Photoemission spectra (EDCs) near these two electron pockets. The EDCs at $\Gamma$ and at the center of the $\beta$ pocket are marked as blue lines. There are sharp EDC peaks for the corner $\beta$ pocket. The observed band is very shallow; the band bottom is only $\sim$5 meV below the Fermi level.  (f) Momentum distribution curves near the gapped region marked by the dashed square in (b) between the binding energy of 50 meV and 29 meV. The MDC at the gap center at a binding energy of 38 meV is marked as red.}


\end{figure*}

\end{document}